\documentclass[twocolumn,prd,floatfix,preprintnumbers,nofootinbib,superscriptaddress]{revtex4-2}

\usepackage{bm}
\usepackage{times}
\usepackage{amssymb,amsbsy,amsmath,amsfonts}
\usepackage{graphicx}
\usepackage{float}
\usepackage{color}
\usepackage{morefloats}
\usepackage{rotating}
\usepackage{srcltx}
\usepackage{slashed}
\usepackage{subfigure}
\usepackage{multirow,diagbox}
\usepackage{verbatim}
\usepackage{tabularx}
\usepackage{tabularray}
\usepackage[outdir=./]{epstopdf}
\usepackage[normalem]{ulem}
\usepackage{hyperref}
\usepackage{tikz}
\hypersetup{colorlinks=true,linkcolor=blue,citecolor=blue,menucolor=black,urlcolor=black,filecolor=blue}

\begin{document}

\title{Studying the $K \Lambda$ strong interaction using femtoscopic correlation functions and the $N^*(1535)$}

\author{Si-Wei Liu}~\email{liusiwei@impcas.ac.cn}
 \affiliation{State Key Laboratory of Heavy Ion Science and Technology, Institute of Modern Physics, Chinese Academy of Sciences, Lanzhou 730000, China} 
\affiliation{School of Nuclear Sciences and Technology, University of Chinese Academy of Sciences, Beijing 101408, China}
\author{Ju-Jun Xie}~\email{xiejujun@impcas.ac.cn}
\affiliation{State Key Laboratory of Heavy Ion Science and Technology, Institute of Modern Physics, Chinese Academy of Sciences, Lanzhou 730000, China} 
\affiliation{School of Nuclear Sciences and Technology, 
    University of Chinese Academy of Sciences, Beijing 101408, China}
\affiliation{Southern Center for Nuclear-Science Theory (SCNT), Institute of Modern Physics, Chinese Academy of Sciences, Huizhou 516000, China}
\date{\today}

\begin{abstract}
    We investigate the $K \Lambda$ strong interaction dynamics around the energy region of $N^*(1535)$ resonance through femtoscopic correlation functions within the Koonin-Pratt formalism, where coupled-channel effects of $\pi N$, $K\Lambda$, $K \Sigma$ and $\eta N$ channels are taken into account, by means of which the $N^*(1535)$ resonance can be dynamically generated in the chiral unitary approach. Two-body scattering amplitudes are calculated using the chiral unitary approach, and the corresponding wave functions and correlation functions are obtained. Our analysis shows that the $\pi N$ channel becomes important to the $K^+\Lambda$ correlation function through coupled channel effects, even if its mass threshold is far from the energy region of the $N^*(1535)$. Furthermore, with the picture that the $N^*(1535)$ is a dynamically generated state, we can get a good reproduction of the experimental data on the $K^+\Lambda$ correlation function. This work establishes femtoscopic correlation functions as a sensitive probe for coupled-channel dynamics and validates the Koonin-Pratt framework in multi-channel hadronic systems, offering critical insights into some exotic hadron states and strong interaction mechanisms.
\end{abstract}

\maketitle

\section{Introduction} \label{sec:Introduction}

The investigation of $K\Lambda$ final state production offers significant opportunities for advancing our understanding about the light baryon resonance spectrum in the medium-energy region. Particularly it is worthy to note that this channel is around the energy of the nucleon excited resonance $N^*(1535)$ has been demonstrated to play a crucial role in numerous processes in hadron physics. This resonance state not only contributes substantially to the dynamics of $K \Lambda$ production but also serves as a key element to understand the complex structure of excited baryon states. With spin-parity quantum numbers $J^P=1/2^-$, the $N^*(1535)$ is conventionally interpreted in the constituent quark model as the first orbital excitation of the nucleon~\cite{ParticleDataGroup:2024cfk}. However, its measured Breit-Wigner mass, $M_{N^*(1535)} = 1530 \pm 15$ MeV, unexpectedly exceeds that of the first radial excitation $N^*(1440)$ with Breit-Wigner mass $M_{N^*(1440)} = 1440 \pm 30$ MeV~\cite{ParticleDataGroup:2024cfk,Zou:2010tc,Zou:2013af,Zou:2018lse}, which is the so-called `mass reversal problem' in the classical quark model. Furthermore, despite its nominal $u$ and $d$ valence quark composition, this resonance exhibits an anomalously strong coupling to the $\eta N$ channel typically associated with strangeness. In the new measurement by the BESIII collaboration~\cite{BESIII:2024vqu}, the ratio of decay widths $\Gamma_{N^*(1535) \to N \eta} / \Gamma_{N^*(1535) \to N\pi}$ is determined to be $0.99 \pm 0.05 \pm  0.19$, which is much larger than the predictions based on flavor symmetry and phase space~\cite{Olbrich:2017fsd}. These distinctive characteristics indicate the presence of a non-conventional hadronic structure beyond the standard $qqq$ configuration in $N^*(1535)$. If we consider $N^*(1535)$ as a meson-baryon dynamically generated state~\cite{Inoue:2001ip,Bruns:2010sv,Khemchandani:2013nma,Garzon:2014ida,Wang:2023snv}, or if we account for significant five-quark components in its wave function~\cite{Liu:2005pm,Zou:2006tw,An:2018vmk,Xu:2020ppr,Zou:2021sha}, the strong coupling of $N^*(1535)$ to the $N \eta$ channel can be naturally explained. Within the above hadronic molecule picture~\footnote{Here, we do not distinguish between the meson-baryon dynamically generated state and the excited baryon state with significant five-quark components.} for the $N^*(1535)$, its couplings to strangeness channels~\footnote{These channels are above the mass threshold of $N^*(1535)$.}, such as $K\Lambda$, $K\Sigma$, $N\eta'$, and $N\phi$, can be also strong~\cite{Inoue:2001ip,Zou:2006tw,Zou:2021sha}. For example, based on its strong couplings to the strangeness channels, the role of $N^*(1535)$ was investigated in the reactions or decays: $pp \to p K^+ \Lambda$~\cite{Zou:2007rx,Xie:2011me}, $J/\psi \to \bar{p} p \eta$~\cite{Geng:2008cv}, $\pi^- p \to n \phi$ and $pp \to pp \phi$~\cite{Xie:2007qt,Doring:2008sv}, $\pi^- p \to \eta n$~\cite{Lu:2013jva}, $\pi^- p \to K^0 \Lambda$~\cite{Wu:2014yca}, $\gamma p \to \phi K^+ \Lambda$~\cite{Lu:2014yba}, $\pi N \to N \eta'$ and $J/\psi \to p\bar{p} \eta'$~\cite{Cao:2014mea}, and $\gamma p \to p \eta \phi$~\cite{Fan:2019lwc}, where these available experimental data can be well reproduced. These above works partly support the hadronic molecule nature of $N^*(1535)$. On the other hand, within the framework of a hadronic molecule picture for the $N^*(1535)$ state, its production in the $\Lambda^+_c \to \bar{K}^0\eta p$ decay has also been evaluated~\cite{Xie:2017erh,Pavao:2018wdf,Xie:2020jlz,Li:2024rqb}.

Even acknowledging the significance of the molecular components in the wave function of the $N^*(1535)$ resonance, many studies suggest that the three-quark components also play a crucial role in the structure of this state~\cite{Hyodo:2008xr,Sekihara:2014kya,Sekihara:2015gvw,Tan:2025kjk}. Under the quenched quark model, the results of Ref.~\cite{Tan:2025kjk} support that the $N^*(1535)$ is a three-quark state. Therefore, more works are still needed to study the nature of the $N^*(1535)$ state. 

Recently, the femtoscopy techniques have emerged as a powerful tool for probing the emission source geometry, two-body hadronic wave functions, and effective hadron-hadron interactions through precision measurements of correlation functions in heavy-ion and $p$-$p$ collisions~\cite{STAR:2014dcy,ALICE:2020ibs,ALICE:2017iga,ALICE:2021njx,ALICE:2021cpv,Sarti:2023wlg,Encarnacion:2024jge}. Since high-quality measurements about the hadron-hadron scattering processes are not suitable for unstable particles in experimental studies, these correlation measurements provide a cleaner experimental environment with reduced resonance contamination compared to conventional approaches, serving as a complementary constraint to scattering experiments and an alternative way to investigate the  invariant mass spectrum for a pair of particles~\cite{Haidenbauer:2018jvl,Liu:2022nec,Liu:2023uly,Liu:2023wfo,Molina:2023oeu,Ikeno:2023ojl,Sarti:2023wlg,Li:2024tof,Feijoo:2024bvn,Yan:2024aap,Ge:2025put}. Femtoscopic correlations between hadron pairs in momentum space have become a powerful tool for studying the hadron-hadron interaction and hadron dynamics, especially for theses exotic hadrons with molecular nature and the near-threshold resonances~\cite{Vidana:2023olz,Albaladejo:2023pzq,Khemchandani:2023xup,Albaladejo:2023wmv,Feijoo:2023sfe,Liu:2024uxn,Liu:2024nac,Abreu:2025jqy,Albaladejo:2025lhn,Li:2024tvo}. In addition, the experimental measurements can be used to determine the free model parameters of the low-energy effective theories.

On the experimental side, the first measurements of the correlation function for three distinct charge combinations of $K^+\Lambda$, $K^-\Lambda$, and $K^0_S\Lambda$ were reported by the ALICE collaboration~\cite{ALICE:2020wvi,ALICE:2023wjz}, enabling us to investigate their strong interactions and extract scattering parameters for all three charge configurations with high precision. On the theoretical side, the interactions of $\eta N$, $K\Lambda$ and $K\Sigma$ were evaluated within the chiral unitary approach, and it was found that the corresponding correlation functions could illuminate the relation between the $N^*(1535)$ state and these coupled channels~\cite{Molina:2023jov,Li:2023pjx}. The theoretical results for the $K\Lambda$ correlation functions of Refs.~\cite{Molina:2023jov,Li:2023pjx} show clear cusps around the mass thresholds of $\eta N$ and $K\Sigma$ channels. However, these cusp structures have not been observed in the available experimental data~\cite{ALICE:2020wvi,ALICE:2023wjz}. It is worthy to mention that the $\pi N$ channel was not taken into account in the works of Refs.~\cite{Molina:2023jov,Li:2023pjx}. As noted in Refs.~\cite{Encarnacion:2024jge,Haidenbauer:2018jvl,Liu:2022nec}, the influence of this channel can significantly affect the shape of the correlation functions. Therefore, the $\pi N$ channel is expected to play a crucial role in the $K^+ \Lambda$ correlation function. Along this line, based on the previous works of Refs.~\cite{Molina:2023jov,Li:2023pjx}, we study the $K \Lambda$ correlation function within the chiral unitary approach with the coupled channels, $\pi N$, $K\Lambda$, $K \Sigma$, and $\eta N$, from where the $N^*(1535)$ is dynamically generated. Taking into account the Koonin-Pratt theoretical framework \cite{Pratt:1984su,Pratt:1990zq,Lednicky:1981su}, we calculate the scattering amplitudes using the chiral unitary approach, from which we subsequently derive the wave functions and momentum correlation functions. It will be demonstrated that the resulting correlation functions are smooth due to the inclusion of the $\pi N$ channel. In addition, the obtained properties of the $N^*(1535)$ state are the same as in the molecular picture.

This article is organized as follows: Sec.~\ref{subsec:Koonin-Pratt-formalism} details the derivation of correlation functions within the Kroonin-Pratt formalism. Sec.~\ref{subsec:Chiral-unitary-approach} briefly introduces the chiral unitary method along with computational formulas for relevant observables. Sec.~\ref{subsec:The effect of coupled channel} decomposes the correlation functions to analyze contributions from individual coupled channels. Sec.~\ref{subsec:Fitting_results_of_CF} presents numerical results for the $K^+\Lambda$ correlation function. The study concludes with a brief summary in Sec.~\ref{sec:Summary and Conclusions}.

\section{Formalism} \label{sec:Formalism}

\subsection{The Koonin-Pratt formalism for the correlation function in coupled channels} \label{subsec:Koonin-Pratt-formalism}

In this section, we will introduce and derive a common formalism of the correlation function. The definition of the correlation function for the $i$-th channel in the coupled channels is given by~\cite{Pratt:1984su,Pratt:1990zq,Lednicky:1981su,Vidana:2023olz},
\begin{equation}
    \begin{aligned}
        C_i(\vec{p})=\int_{0}^{\infty} d^3\vec{r}~S(\vec{r})\left | \Psi_i(\vec{r},\vec{p})\right |^2,
        \label{equ:cf-defination}
    \end{aligned}
\end{equation}
where the $\Psi_i(\vec{r},\vec{p})$ is the two-body scattering wave function of the $i$-th hadron pair in coupled channels and $\vec{p}$ is the three momentum in the pair rest frame. $S(\vec{r})$ is the emitting source function that is usually parameterized as a normalized Gaussian function,
\begin{equation}
    \begin{aligned}
        S\left(\vec{r}\right)=\frac{e^{-\frac{\left | \vec{r} \right |^2}{4R^2}}}{\left(4\pi R^2\right)^{3/2}},
    \end{aligned}
\end{equation}
where $R$ provides the range of the production source. Its typical value for $pp$ collisions is about 1 fm. However, as the feed-down contributions mentioned in Sec.~\ref{subsec:Fitting_results_of_CF}, there may be some long-lived resonances in the emitting progress, which will cause the particles to be emitted from farther apart. It means that the size of emitting source will be incresed and an exponential tail appears in the source distribution~\cite{ALICE:2020ibs,ALICE:2023wjz}. Therefore, inspired by this picture, we need to use a Gaussian core and a non-Gaussian halo to build source function as in Ref.~\cite{ALICE:2023wjz},
\begin{equation}
    \begin{aligned}
        S_{\text{eff}}(\vec{r})=\lambda_S\left[\omega_S S(\vec{r},R_1)+(1-\omega_S)S(\vec{r},R_2)\right].
        \label{equ:two-Gaussians}
    \end{aligned}
\end{equation}
Here we take $R_1=1.202^{+0.043}_{-0.042}~\text{fm}$, $R_2=2.330^{+0.050}_{-0.045}~\text{fm}$, $\lambda_S=0.9806^{+0.0006}_{-0.0008}$, and $\omega_S=0.7993^{+0.0037}_{-0.0027}$ as used in Ref.~\cite{ALICE:2023wjz}. 

Furthermore, the two-body scattering wave function after the final state interaction can be obtained as,
\begin{eqnarray}
        \Psi_{ij}(\vec{r},\vec{p}) &=& \delta_{ij}\Phi(\vec{r},\vec{p})+\Psi'_{ij}(\vec{r},\vec{p}) \nonumber \\
        &=& \delta_{ij}e^{i \vec{p} \cdot \vec{r}}+\Psi'_{ij}(\vec{r},\vec{p}),
        \label{equ:scattering-wave-function-defination}
    \end{eqnarray}
which consists of a plane wave $e^{i \vec{p} \cdot \vec{r}}$ and a scattered wave $\Psi'_{ij}(\vec{r},\vec{p})$. Therefore, considering all coupled channels, the total two-body wave function can be obtained,
\begin{equation}
    \begin{aligned}
       & \left | \Psi_i(\vec{r},\vec{p})\right |^2 = \sum_j w_{j}\left | \Psi_{ij}(\vec{r},\vec{p})\right |^2 \\
        &= w_{i}\left | \Psi'_{ii}(\vec{r},\vec{p})+e^{i \vec{p} \cdot \vec{r}}\right |^2+\sum_{j\neq i}w_{j} \left | \Psi'_{ij}(\vec{r},\vec{p}) \right |^2,
        \label{equ:two-body wave function square}
    \end{aligned}
\end{equation}
where $w_{j}$ represents the production ratio of $j$-th coupled channel particles generated in collisions to the number of $i$-th channel. Due to the lack of experimental information for $w_{j}$, we set all the $w_{j}$ to 1 and their corresponding emitting source size $R$ to the same value for simplification~\cite{Vidana:2023olz,Molina:2023oeu,Molina:2023jov}. Since the plane wave $e^{i \vec{p} \cdot \vec{r}}$ can be expanded to Legendre polynomials,
\begin{equation}
    \begin{aligned}
        e^{i \vec{p} \cdot \vec{r}}=e^{i |\vec{p}||\vec{r}|\cos \theta}=\sum_{l=0}^{\infty}i^l (2l+1) j_l(|\vec{p}||\vec{r}|) P_l(\cos \theta),
    \end{aligned}
\end{equation}
where $j_l(x)$ is the $l$-th order spherical Bessel function, and because the value of this integral $\int_{-1}^{1} d\cos\theta~P_l(\cos \theta)$ is non-zero only if $l=0$, the following equation can be obtained,
\begin{equation}
    \begin{aligned}
        \int_{-1}^{1} d\cos \theta~e^{i \vec{p} \cdot \vec{r}}=\int_{-1}^{1} d\cos \theta~j_0(|\vec{p}||\vec{r}|),
        \label{equ:j0}
    \end{aligned}
\end{equation}
where $j_0(x) = {\rm sin} x/x$. Therefore, combining the Eq.~\eqref{equ:two-body wave function square} and Eq.~\eqref{equ:j0}, the correlation function $C_i(\vec{p})$ in Eq.~\eqref{equ:cf-defination} can be simplified as following,
\begin{widetext}
\begin{eqnarray}
C_i(\vec{p}) = 1+\int_{0}^{\infty} d^3\vec{r}~S_{12}(\vec{r}) \left (2\text{Re}\left[\Psi^{'}_{ii}(\vec{r},\vec{p})j_0\left(\left |\vec{p}\right | \left |\vec{r}\right |\right)\right]+\sum_{j} \left | \Psi'_{ij}(\vec{r},\vec{p}) \right |^2\right ).
        \label{equ:cf-results}
\end{eqnarray}
\end{widetext}

Following the derivations of the Refs.~\cite{Vidana:2023olz,Albaladejo:2023wmv}, the analytic expression of correlation function can be obtained in the Koonin-Pratt formalism. First, the two-body scattering wave function $\Psi_{ij}(\vec{r},\vec{p})$ in Eq.~\eqref{equ:scattering-wave-function-defination} can be derived from Lippmann-Schwinger equation,
\begin{widetext}
\begin{equation}
    \begin{aligned}
        \Psi_{ij}(\vec{r},\vec{p})&=\delta_{ij}e^{i \vec{p}\vec{r}}+T_{ij}(\vec{r},\vec{p})\int_{0}^{\infty} \frac{d^4q}{(2\pi)^4} \frac{i (2M_j)^n e^{i\vec{q}\vec{r}}}{(q^2-M_j^2+i\epsilon)((P-q)^2-m_j^2+i\epsilon)}=\delta_{ij}e^{i \vec{p}\vec{r}}+T_{ij}(\vec{r},\vec{p})\tilde{G}_j(\vec{r},\vec{p}),
        \label{equ:scattering-wave-function-KP-results}
    \end{aligned}
\end{equation}
where
\begin{equation}
  \begin{aligned}
   & \tilde{G}_j(\vec{r},\vec{p}) = (2M_j)^n\int_{0}^{\infty}\frac{d^3\vec{q}}{(2\pi)^3}  \frac{(\omega_{M_j}+\omega_{m_j})j_0(\left |\vec{q}\right |\left |\vec{r}\right |)}{2\omega_{M_j} \omega_{m_j}\left[s-(\omega_{M_j}+\omega_{m_j})^2+i\epsilon\right]} =(2M_j)^n \times \\ & \int_{0}^{\infty}\frac{d^3\vec{q}}{(2\pi)^3}  \left[\frac{(\omega_{M_j}+\omega_{m_j})j_0(\left |\vec{q}\right |\left |\vec{r}\right |)-j_0(\left |\vec{q}_{j,\text{on}}\right |\left |\vec{r}\right |)}{2\omega_{M_j} \omega_{m_j}\left[s-(\omega_{M_j}+\omega_{m_j})^2+i\epsilon\right]} + j_0(\left |\vec{q}_{j,\text{on}}\right |\left |\vec{r}\right |)  \frac{\omega_{M_j}+\omega_{m_j}}{2\omega_{M_j} \omega_{m_j}\left[s-(\omega_{M_j}+\omega_{m_j})^2+i\epsilon\right]}\right],
      \label{equ:tilde-G}
  \end{aligned}
\end{equation}
\end{widetext}
where $\left|\vec{q}_{j,\text{on}}\right|=\lambda^{1/2}(s,M_j^2,m_j^2)/(2\sqrt{s})$, $\omega_{M_j}=\sqrt{M_j^2+|\vec{q}|^2}$, and $s=P^2$ are the on-shell three momenta, the energy, and the invariant mass square of the $j$-th hadron pair, respectively. And $\lambda(a,b,c)=a^2+b^2+c^2-2ab-2ac-2bc$. $M_j$ represents the fermionic mass in $j$-th coupled channel which is just a convention due to fermion, where $n$ is the number of fermions in loop function.  The $T_{ij}$ matrix is the two-body scattering amplitude that can be calculated by the chiral unitary approach.

Finally, the analytic equation for the correlation function $C_i(\vec{p})$ in the Koonin-Pratt formalism can be easily obtained,
\begin{eqnarray}
        C_i(\vec{p}) &=& 1+\int_{0}^{\infty} d^3\vec{r}~S_{12}(\vec{r})\left (\sum_{j} \left | T_{ij}(\vec{r},\vec{p})\tilde{G}_j(\vec{r},\vec{p}) \right |^2 \right. \nonumber \\
      && \left.+2\text{Re}\left[T_{ii}(\vec{r},\vec{p})\tilde{G}_i(\vec{r},\vec{p})j_0\left(\left |\vec{p}\right | \left |\vec{r}\right |\right)\right]\right ).
      \label{equ:cf-KP-results}
\end{eqnarray}

\subsection{The chiral unitary approach for the $K\Lambda$ interaction within coupled channels} \label{subsec:Chiral-unitary-approach}

Within the chiral unitary approach, the unitary two-body scattering matrix $T$ among the strangeness $S=0$ and charge$=1$ states $K^0 \Sigma^+$, $K^+ \Sigma^0$, $K^+ \Lambda$, $\pi^+ n$, $\pi^0 p$, and $\eta p$ can be obtained with the Bethe-Salpeter equation~\cite{Molina:2023uko,Duan:2024ygq,Khemchandani:2016ftn},
\begin{equation}
    \begin{aligned}
        T=(1-VG)^{-1}V,
        \label{equ:Tmatrix}
    \end{aligned}
\end{equation}
where $V$ is the pseudoscalar meson-baryon interaction potential matrix obtained from the leading order chiral Lagrangian, as follows
\begin{equation}
    \begin{aligned}
        V_{ij} & =  -\frac{C_{ij}}{4f_i f_j}(2\sqrt{s}-M_i-M_j)\sqrt{\frac{(M_i+E_i)(M_j+E_j)}{4M_iM_j}},
        \label{equ:V-formalism}
    \end{aligned}
\end{equation}
where $f_j$, $M_j$, $k_j$, and $E_j$ ($j=$ $K^0 \Sigma^+$, $K^+ \Sigma^0$, $K^+\Lambda$, $\pi^+ n$, $\pi^0 p$, or $\eta p$)  are the meson decay constant, baryon mass, meson energy, and baryon energy of the $j$-th channel, respectively. Here we take $f_\pi=93$ MeV, $f_K=1.22f_\pi$ and $f_\eta=1.3f_\pi$, $m_{K^+} = 493.68$, $m_{K^0} = 497.61$, $m_{\pi^+} = 139.57$, $m_{\pi^0} = 134.98$, $m_\eta = 547.86$ MeV, and $M_{p} = 938.27$, $M_{n} = 939.57$, $M_{\Sigma^+} = 1189.37$, $M_{\Sigma^0} = 1192.64$, and $M_{\Lambda} =1115.68$ MeV. The factor $C_{ij}$ is symmetric and we take their values as used in Refs.~\cite{Inoue:2001ip,Molina:2023jov}. Note that we have included the $\pi^+n$ and $\pi^0 p$ channels, although their mass thresholds are far from the energy region of $N^*(1535)$. 

Moreover, a critical distinction arises in the treatment of this loop function integrals, which exhibit logarithmic divergences requiring cut-off~\cite{Zhu:2022guw,Aceti:2015zva,Wang:2020pem} or dimensional regularization~\cite{Nishibuchi:2023acl,Feijoo:2023wua}. The analytic form of the loop function within the dimensional regularization is as follows:
\begin{eqnarray}
        G_j(s) \!\! &=& \!\! \frac{2M_j}{16\pi^2}\Bigg\{\alpha_j(\mu) + \text{ln}\frac{M_j^2}{\mu^2}+\frac{s+\Delta}{2s}\text{ln}\frac{m_j^2}{M_j^2} + \nonumber  \\
     && \!\! \!\! \frac{p_j}{\sqrt{s}}\bigg[\text{ln}\left(s-\Delta+2p_j\sqrt{s}\right)  -\text{ln}\left(-s-\Delta+2p_j\sqrt{s}\right) +  \nonumber \\
   && \!\! \!\! \!\!  \text{ln}\left(s+\Delta+2p_j\sqrt{s}\right)-\text{ln}\left(-s+\Delta+2p_j\sqrt{s}\right)\bigg]\Bigg\},
 \label{equ:G-dimension}
    \end{eqnarray}
with $\Delta=m_j^2-M_j^2$, $p_j=\lambda^{1/2}(s,M_j^2,m_j^2)/(2\sqrt{s})$ and $\lambda(a,b,c)=a^2+b^2+c^2-2ab-2ac-2bc$. While $\mu$ is the scale of dimensional regularization and $\alpha_j(\mu)$ is the subtraction constant of $j$-th channel. In this paper, we take $\mu=1200$ MeV for all the coupled channels and $\alpha_j(\mu)$~\footnote{To minimize free parameters, we will take $\alpha_{K^+ \Sigma^0} = \alpha_{K^0 \Sigma^+} \equiv \alpha_{K\Sigma}$ and $\alpha_{\pi^+ n} = \alpha_{\pi^0 p } \equiv \alpha_{\pi N}$.} are free parameters which are determined by fitting them to the experimental results of the $K \Lambda$ correlation function and the pole position of $N^*(1535)$ state.

With the obtained two-body scattering amplitudes, $T_{ij}$, one can calculate the scattering length and the effective range, which can be obtained as~\cite{Inoue:2001ip,Molina:2023jov,Li:2023pjx},
\begin{equation}
    \begin{aligned}
        a_j&=\left.\frac{2M_jT_{jj}}{8\pi\sqrt{s}}\right|_{s=s_{th,j}},\\
        r_{j}&=2\frac{\partial}{\partial p^2}\left.\left[-\frac{8\pi\sqrt{s}}{2M_jT_{jj}}+ip_j\right]\right|_{s=s_{th,j}},
    \end{aligned}
\end{equation}
where $s_{th,j}$ is the threshold energy square of the $j$-th channel.

In order to obtain the pole position of $N^*(1535)$ which is dynamically generated by the final state interaction, we need to extrapolate the energy to the complex plane. The analytical continuation of the loop function is given by:
\begin{equation}
    \begin{aligned}
    G_{j}(s) & = \left\{\begin{array}{ll}
        G_{j}^{(I)}(s), & \text { for } \text{Re}\{\sqrt{s}\}<\left(m_{j}+M_{j}\right) \\
        G_{j}^{(I I)}(s), & \text { for } \text{Re}\{\sqrt{s}\} \geq\left(m_{j}+M_{j}\right)
    \end{array},\right.
    \label{equ:G-complex-plane}
    \end{aligned}
\end{equation}
where $G_{j}^{(I)}(s)$ is the loop function as in Eq.~\eqref{equ:G-dimension} and 
\begin{equation}
    \begin{aligned}
    G_{j}^{(I I)}(s)=G_{j}^{(I)}(s)+i\frac{2M_j}{4\pi\sqrt{s}}p_j.
    \label{equ:G-second-riemann}
    \end{aligned}
\end{equation}

Then the pole position of $N^*(1535)$ satisfies $\left.\text{det}\left(1-VG\right)\right|_{s=s_{pole}}=0$ with $\sqrt{s_{pole}} = M_{N^*(1535)} - i \Gamma_{N^*{(1535)}}/2$. Once the pole position is determined, the coupling of $N^*(1535)$ resonance to each channel, $g_i$, can be obtained as,
\begin{equation}
    \begin{aligned}
    g_ig_j= \left. \frac{r}{2\pi}\int^{2\pi}_0 T_{ij}(\sqrt{s})e^{i\theta}d\theta \right |_{r \to 0},
    \end{aligned}
\end{equation}
with $\sqrt{s} = \sqrt{s_{pole}} + r e^{i\theta}$. In general, the coupling constant $g_j$ is complex.

With the obtained coupling constant, the partial decay width of the $N^*(1535)$ can be calculated by,
\begin{eqnarray}
    \Gamma_j &=& \frac{|g_j|^2}{4\pi} \frac{(M_{N^*(1535)}^2 + M_j^2 - m^2_j)}{4 M^2_{N^*(1535)}} \times \nonumber \\
    && \lambda^{1/2}(M^2_{N^*(1535)},M^2_j,m^2_j).
    \end{eqnarray}

\section{Results and Discussions} \label{sec:Results}
\subsection{The $\pi N$ channel effects in the $K^+\Lambda$ correlation function} \label{subsec:The effect of coupled channel}

\begin{figure}[htbp]
    \includegraphics[scale=0.5]{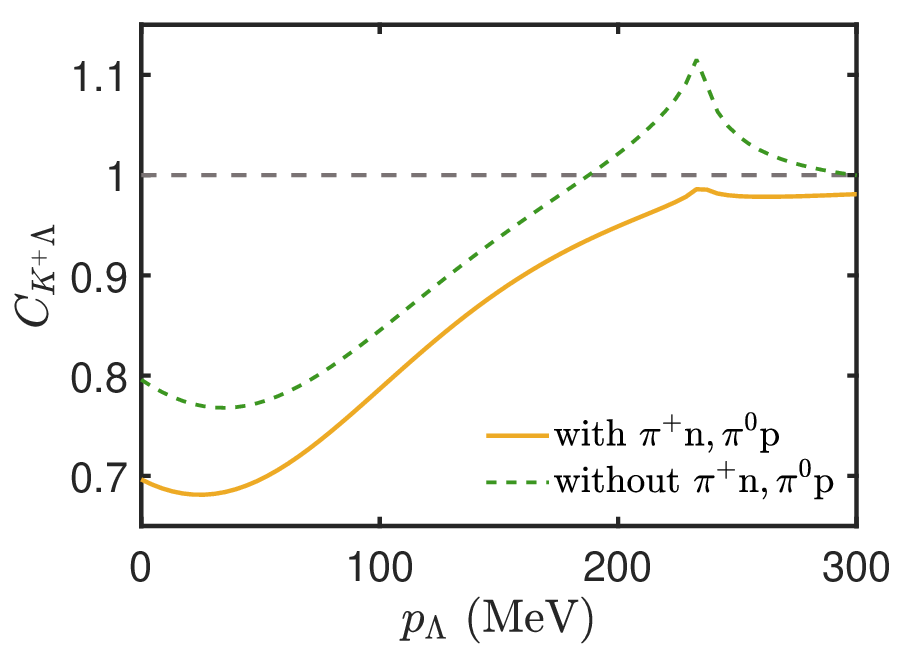}
    \caption{The contributions of each coupled channel to the $K^+\Lambda$ correlation function. The dotted green line is the result without $\pi^+n$ and $\pi^0p$ channels, and the solid orange line is the result with $\pi^+n$ and $\pi^0p$. The dashed curve stands for unity for comparison.}
    \label{fig:coupled_effect}
\end{figure}

With the two-body scattering amplitudes $T$ obtained within the chiral unitrary approach, one can easily obtain the $K^+ \Lambda$ correlation function as in Eq.~\eqref{equ:cf-KP-results}. We show the numerical results as a function of $p_\Lambda$, the relative momentum of the $\Lambda$ in the $K^+\Lambda$ system, in Fig.~\ref{fig:coupled_effect}, where the green-dashed curve stands for the results without including the $\pi^+n$ and $\pi^0p$ channels, and the orange-solid curve stands for the results considering the $\pi^+n$ and $\pi^0p$ channels. To investigate the significance of the $\pi N$ channel, we adopt the model parameters from Refs.~\cite{Molina:2023jov,Li:2023pjx}, which utilize a cut-off regularization scheme for the meson-baryon loop functions with $q_{\rm max}$ = 630 MeV. This specific choice of regularization scheme and parameter value ensures consistency with previous theoretical studies while providing a robust basis for evaluating the contributions of the $\pi^+n$ and $\pi^0p$ channels.

From Fig.~\ref{fig:coupled_effect}, one can see that, without considering the $\pi^+n$ and $\pi^0p$ channels, the numerical results show a clear threshold cusp at $K \Sigma$ mass threshold as pointed out in Refs.~\cite{Molina:2023jov,Li:2023pjx}, this is because of the singularity in the modified propagator $\tilde{G}$'s denominator [see Eq.~\eqref{equ:tilde-G}]. With the inclusion of $\pi^+n$ and $\pi^0p$ channels, the peak magnitudes of the $K^+ \Lambda$ correlation function are reduced. In the whole energy region we considered, the obtained $K^+\Lambda$ correlation function is smaller than one. This indicates that the introduction of lower-energy $\pi^+n$ and $\pi^0p$ channels are important to the scattering amplitude through coupled-channel interactions, even if the threshold of the $\pi N$ channel is far from the considered energy region. These findings qualitatively validate this theoretical formalism's predictive power for multi-channel hadronic systems, establishing femtoscopic correlation functions as sensitive probes of coupled-channel effects.

\subsection{Fitting results for the $K^+ \Lambda$ correlation function}~\label{subsec:Fitting_results_of_CF}

\begin{figure*}[htbp]
    \centering
    \includegraphics[scale=0.8]{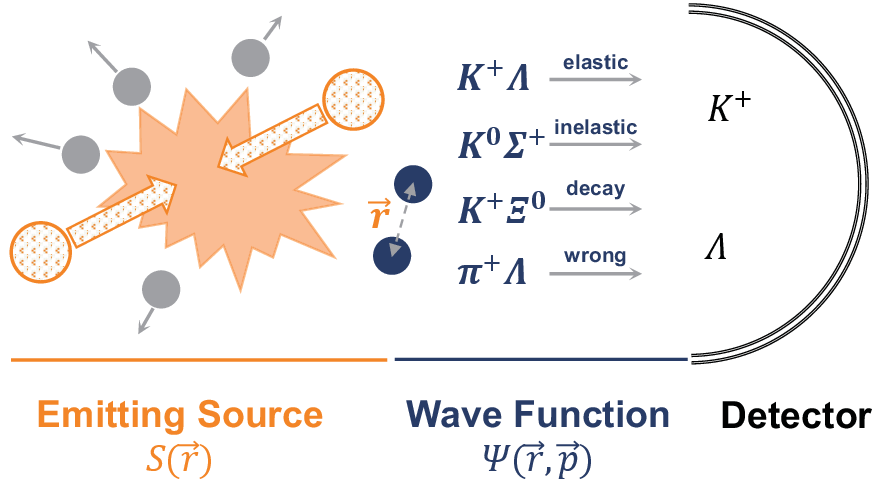}
    \caption{The schematic diagram for the $K^+ \Lambda$ final state interaction in the relativistic heavy-ion collision.}
    \label{fig:schematic_diagram}
\end{figure*}

We firstly introduce the process of the $K^+ \Lambda$ final state interaction in the relativistic heavy-ion collision~\cite{ALICE:2023wjz,ALICE:2020wvi,ALICE:2018ysd,ALICE:2019eol} which is schematically illustrated in Fig.~\ref{fig:schematic_diagram}. The production process can be described through two distinct stages: (1) In the initial phase, kaons, pions, and hyperons are generated from the heavy-ion collision at extremely high energies, forming what is referred to as emitting source [$S(\vec{r})$]. (2) Subsequently, through final-state interactions, detectable hadron pairs $K^+$ and $\Lambda$ were left and ultimately measured by the detector system. In this case, the $K^+ \Lambda$ correlation function can be measured at low relative momenta.

Note that there are three distinct contributions involved in the measured $K^+\Lambda$ correlation functions: First, the genuine contributions arising from $K^+\Lambda$ pairs interacting via coupled-channel elastic and inelastic interactions (e.g., $K^0\Sigma^+, K^+\Lambda \to K^+\Lambda$). Second, the feed-down contributions from intermediate particle decay (e.g., $\Xi^0 \to \Lambda$). Third, the experimental artifacts from particle misidentification. In the following discussion, we only introduce the theoretical calculation process for the genuine contribution, while the other contributions need to be determined through fitting experimental data or direct experimental measurements.

Therefore, to compare with the experimental measurements, the fitting function is defined as  
\begin{equation}
    C_{\text{fit}}=N'\times C_{\text{tot}}\times C_{\text{background}},
\end{equation}
where $N'$ is the global normalization factor allowed to vary freely. The term $C_{\text{background}}$ accounts for residual experimental backgrounds, including hard scattering processes in collisions, unconsidered resonances in $C_{\text{tot}}$, and other effects, as reported in Ref.~\cite{ALICE:2023wjz}. The theoretical component $C_{\text{tot}}$ is expressed as  
\begin{equation}
    C_{\text{tot}}=1+\lambda_{\text{gen}} \left(C_{\text{gen}}-1\right)+\sum_{n}\lambda_n\left(C_n-1\right),
\end{equation}
which includes all detectable $K^+\Lambda$ production mechanisms as mentioned above. The constant $C_{\text{gen}}$ corresponds to the correlation function obtained in Eq.~\eqref{equ:cf-KP-results} and $C_n$ represents the non-genuine contribution. Every contribution is weighted by a normalized parameter $\lambda$ (i.e., $\lambda_{\text{gen}}+\sum_n\lambda_n=1$). Given limited information about $\lambda$ and the absence of experimental analysis of the decaying processes, we make a approximation that all the non-genuine contributions are as a flat constant term $X$ (i.e., $\sum_n\lambda_nC_n=X$). Then the simplified fitting function becomes:  
\begin{equation}
    C_{\text{fit}}=N \times \left(\lambda C_{\text{gen}}+1-\lambda\right)\times C_{\text{background}} ,
    \label{equ:fitting-CF}
\end{equation}
where $N=N'\left(\lambda_{\text{gen}}+X\right)$, which is a normalization global factor close to one. $\lambda=\frac{\lambda_{\text{gen}}}{\lambda_{\text{gen}}+X}$ is a free parameter. In the fitting to the experimental measurements about the $K^+ \Lambda$ correlation function, we take the following two fitting scenarios:

Scenario I: Subtraction constants $a(\mu)$ are fixed as in Ref.~\cite{Inoue:2001ip}, while $N$ and $\lambda$ are free parameters. In this case, there are a total of 28 data points and two free parameters;  

Scenario II: $a_j(\mu)$, $N$, and $\lambda$ are fitted to the $K^+\Lambda$ correlation function. In addition, the $N^*(1535)$ pole position constrained to be $M_{N^*(1535)} = 1510 \pm 10$ MeV and $\Gamma_{N^*(1535)} = 105 \pm 25$ MeV, as quoted in the particle data group (PDG)~\cite{ParticleDataGroup:2024cfk}. In this case, there are 28 data points in total and six free parameters.

\begin{table}[htbp]
\centering
\caption{The fitted model parameters. Note that the subtraction constants for Scenario I were fixed as in Ref.~\cite{Inoue:2001ip}.}
\label{tab:fitting_parameters}
\begin{tblr}{
  width = \linewidth,
  colspec = {Q[227]Q[238]Q[238]},
  cells = {c},
  hline{1-2,9} = {-}{},
}
Scenario                & I                        & II                        \\
$N$                     & $0.99554\pm 0.00045$     & $0.99691\pm 0.00082$      \\
$\lambda$               & $0.87\pm 0.01$           & $0.99\pm 0.09$            \\
$\alpha_{K\Sigma}$           & -2.8                     & $-3.25\pm 0.17$           \\
$\alpha_{K\Lambda}$          & 1.6                      & $1.65\pm 0.35$            \\
$\alpha_{\eta N}$            & 0.2                      & $0.95\pm 0.28$            \\
$\alpha_{\pi N}$             & 2.0                      & $1.92\pm 0.42$            \\
$\chi^2/\rm d.o.f.$     & 1.9                      & 1.1                                       
\end{tblr}
\end{table}

The fitted parameters and the corresponding obtained properties of $N^*(1535)$ with the central values of the fitted model parameters are summarized in Table~\ref{tab:fitting_parameters} and Table~\ref{tab:fitting_properties}, respectively. Note that the obtained $\chi^2/{\rm d.o.f.} = 1.9$ for Scenario I is reasonable small. For Scenario II, the obtained $\chi^2/{\rm d.o.f.}$ is 1.1, which is smaller than the one obtained in Scenario I. And, the fitted model parameters $N$, $\lambda$, $a_{K \Lambda}$, and $a_{\pi N}$ are similar with those in Scenario I within uncertainties. However, the values of $a_{K \Sigma}$ and $a_{\eta N}$ are much different for Scenario II and I.

Based on the obtained properties of the $N^*(1535)$ state shown in Table~\ref{tab:fitting_properties}, we could conclude that both Scenario I and Scenario II can dynamically generate the $N^*(1535)$ state in the chiral unitary approach, and they also give very similar results for other properties of the resonance. In Scenario I, the mass of $N^*(1535)$ resonance, $M_{N^*(1535)} = 1535$ MeV, is heavier than the one, $M_{N^*(1535)} = 1505$ MeV, obtained in Scenario II. In the first case it is closer to the Breit-Wigner mass of $M_{N^*(1535)} = 1530 \pm 15$ MeV~\cite{ParticleDataGroup:2024cfk}.

\begin{table}[htbp]
\centering
\caption{The obtained properties for the $N^*(1535)$ resonance.}
\label{tab:fitting_properties}
\begin{tblr}{
  width = \linewidth,
  colspec = {Q[180]Q[150]Q[212]Q[212]},
  cells = {c},
  cell{1}{1} = {c=2}{0.308\linewidth},
  cell{2}{1} = {c=2}{0.308\linewidth},
  cell{3}{1} = {r=6}{},
  cell{9}{1} = {r=3}{},
  cell{12}{1} = {r=6}{},
  cell{18}{1} = {r=6}{},
  cell{24}{1} = {r=7}{},
  hline{1-3,9,12,18,24,31} = {-}{},
}
Scenario             &                     & I                       & II                        \\
$\sqrt{s_{\rm pole}}$~ (MeV) &                  & $1531 - \text{i}~36.5$& $1505 - \text{i}~40.1$ \\
Coupling             & $g_{K^0\Sigma^+}$   & $2.21-\text{i}~0.18$    & $2.06-\text{i}~0.43$     \\
                     & $g_{K^+\Sigma^0}$   & $1.56-\text{i}~0.13$    & $1.45-\text{i}~0.31$     \\
                     & $g_{K^+\Lambda}$    & $1.37-\text{i}~0.10$    & $0.99-\text{i}~0.11$     \\
                     & $g_{\pi^+ n}$       & $0.55+\text{i}~0.32$    & $0.55+\text{i}~0.25$     \\
                     & $g_{\pi^0 p}$       & $0.40+\text{i}~0.22$    & $0.40+\text{i}~0.17$     \\
                     & $g_{\eta p}$        & $-1.47+\text{i}~0.43$   & $-1.73+\text{i}~0.75$    \\
{Partial Width\\(MeV)}     & $\Gamma_{\pi^+ n}$  & 19.7                   & 17.3                    \\
                     & $\Gamma_{\pi^0 p}$  & 9.9                    & 8.8                     \\
                     & $\Gamma_{\eta p}$   & 41.2                   & 41.4                    \\
{Sacttering Length\\(MeV$^{-1}$)} & $a_{K^0\Sigma^+}$   & $0.28-\text{i}~0.14$    & $0.24-\text{i}~0.12$     \\
                     & $a_{K^+\Sigma^0}$   & $0.19-\text{i}~0.13$    & $0.17-\text{i}~0.11$     \\
                     & $a_{K^+\Lambda}$    & $0.33-\text{i}~0.14$    & $0.26-\text{i}~0.15$     \\
                     & $a_{\pi^+ n}$       & $-0.08$                 & $-0.08$                  \\
                     & $a_{\pi^0 p}$       & $0.02$                  & $0.02$                   \\
                     & $a_{\eta p}$        & $-0.25-\text{i}~0.17$   & $-0.76-\text{i}~0.33$    \\
{Effective Range\\(MeV$^{-1}$)}   & $r_{K^0\Sigma^+}$   & $1.05-\text{i}~1.55$    & $0.71-\text{i}~1.20$     \\
                     & $r_{K^+\Sigma^0}$   & $-1.46-\text{i}~3.69$   & $-1.29-\text{i}~2.68$    \\
                     & $r_{K^+\Lambda}$    & $-0.52-\text{i}~0.38$   & $0.68-\text{i}~1.03$     \\
                     & $r_{\pi^+ n}$       & $-13.39-\text{i}~18.04$ & $-13.75-\text{i}~17.24$  \\
                     & $r_{\pi^0 p}$       & $30.34$                 & $18.48$                  \\
                     & $r_{\eta p}$        & $-8.60+\text{i}~6.77$   & $-3.85+\text{i}~1.52$ 
\end{tblr}
\end{table}

It is worthy to mention that the ratio of the coupling constants $|g_{K^0\Sigma^+}/g_{K^+\Sigma^0}| \approx |g_{\pi^+ n}/g_{\pi^0 p}| \approx \sqrt{2}$ satisfies the isospin $I=1/2$ decomposition~\footnote{We have followed: \begin{eqnarray} |\frac{1}{2}\frac{1}{2}> = \sqrt{\frac{2}{3}} |11> |\frac{1}{2} -\frac{1}{2}> - \sqrt{\frac{1}{3}} |10> |\frac{1}{2} \frac{1}{2}>, \nonumber
\end{eqnarray} with the basis of $|II_Z>$.}, which support that the isospin of $N^*(1535)$ is 1/2. With the obtained total width [$\Gamma_{N^*(1535)} = -2 {\rm Im}(\sqrt{s_{\rm pole}})$] of $N^*(1535)$ and its partial decay widths to the $\pi N$ and $\eta N$ channels, one could easily get the branching fractions,$\\$
Scenario ~ I:
\begin{eqnarray}
Br[N^*(1535) \to \pi N] & = & \frac{\Gamma_{\pi^+ n} + \Gamma_{\pi^0 p}}{\Gamma_{N^*(1535)}} = 41 \%, \\
Br[N^*(1535) \to \eta N] & = & \frac{\Gamma_{\eta p}}{\Gamma_{N^*(1535)}} = 56 \%,
\end{eqnarray}
Scenario ~ II:
\begin{eqnarray}
Br[N^*(1535) \to \pi N] & = & \frac{\Gamma_{\pi^+ n} + \Gamma_{\pi^0 p}}{\Gamma_{N^*(1535)}} = 33 \%, \\
Br[N^*(1535) \to \eta N] & = & \frac{\Gamma_{\eta p}}{\Gamma_{N^*(1535)}} = 52 \%,
\end{eqnarray}
which are in agreement with these values, $Br[N^*(1535] \to \pi N] = (32-52)\% $ and $Br[N^*(1535] \to \eta N] = (30-55)\%$, as quoted in the PDG~\cite{ParticleDataGroup:2024cfk} within uncertainties. In addition, the obtained ratio of decay widths $\Gamma_{N^*(1535) \to N \eta} / \Gamma_{N^*(1535) \to N\pi} = \Gamma_{\eta p} / (\Gamma_{\pi^+ n} + \Gamma_{\pi^0 p})$ is $1.4$ and $1.6$ for Scenario I and II, respectively. Both of the above values are larger than one.

\begin{figure}[htbp]
    \centering
    \includegraphics[scale=0.35]{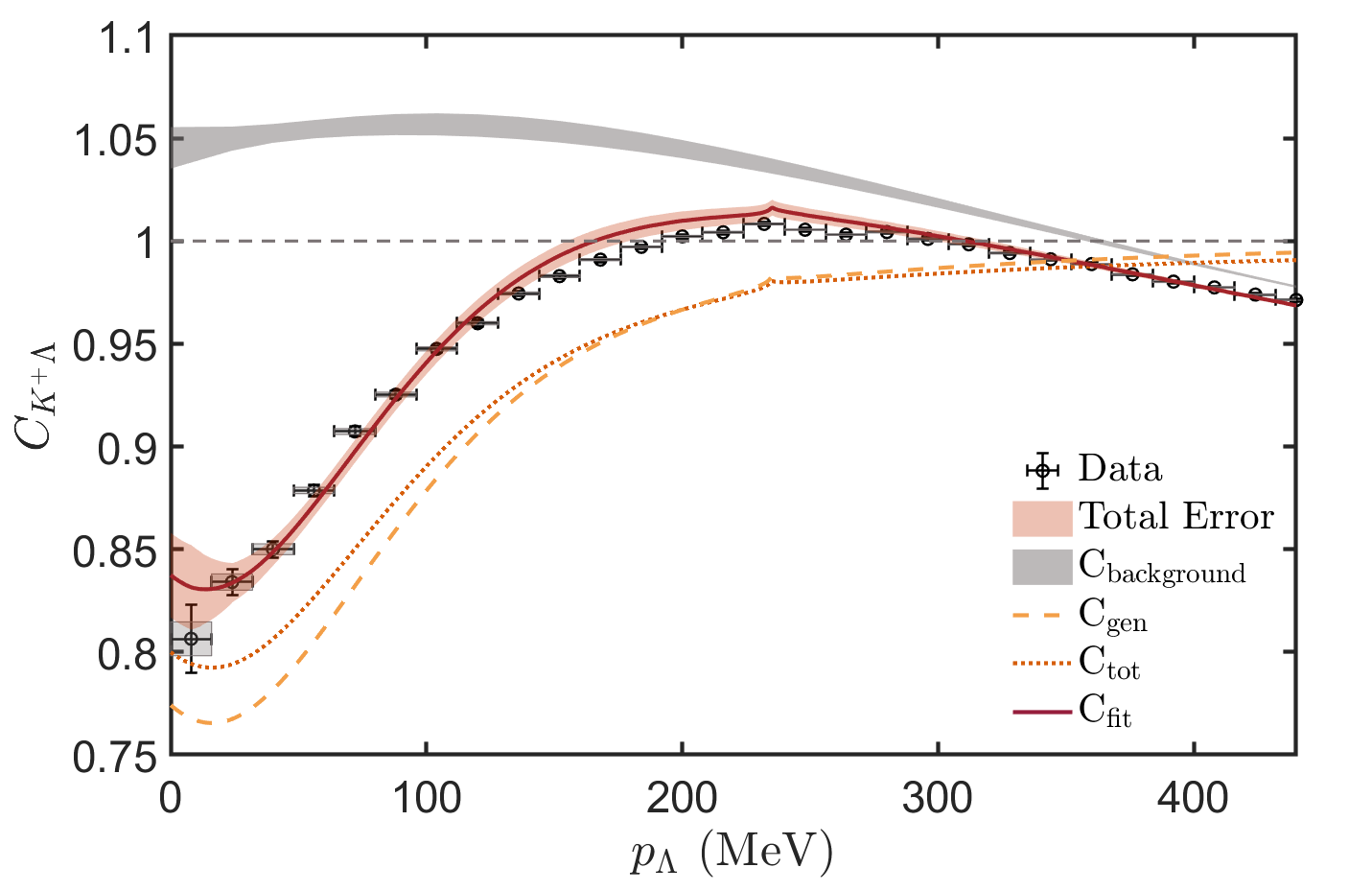}
     \includegraphics[scale=0.35]{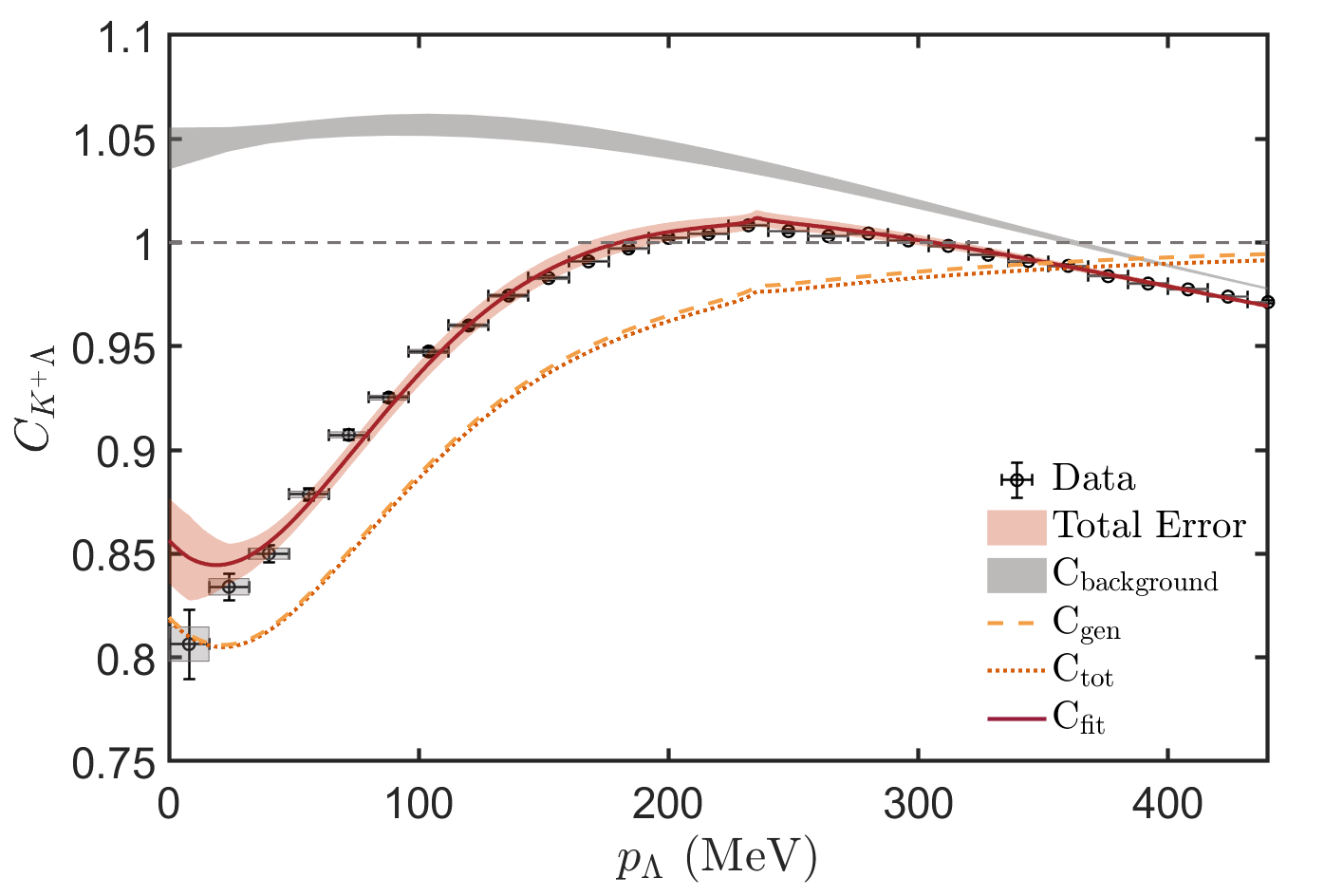}
    \caption{The fitted results of the $K^+\Lambda$ correlation functions for scenario I (up) and II (down). The experimental data are taken from Ref.~\cite{ALICE:2023wjz}.}    \label{fig:Scenario_1}
\end{figure}

Then, we show the fitted numerical results for $K^+ \Lambda$ correlation functions in both scenarios in Fig.~\ref{fig:Scenario_1}, from where one can see that we can get a fairly well reproduction of the experimental measurements about the $K^+ \Lambda$ correlation function within uncertainties. In Fig.~\ref{fig:Scenario_1}, the gray shaded area indicates the experimental background uncertainty $C_{\text{background}}$ (more details can be found in Ref.~\cite{ALICE:2023wjz}). The yellow dashed line represents the theoretical $K^+ \Lambda$ genuine contribution $C_{\text{gen}}$, while the brown dotted line represents the theoretical total contribution $C_{\text{tot}}$ that further incorporates the feed-down and particle misidentification contributions. The red line and the pink shaded area represent the theoretical fitted results $C_{\text{fit}}$ including both the experimental background and experimental uncertainties.

For $p_\Lambda < 200$ MeV, the $K^+\Lambda$ correlation function is smaller than one, and it decreases with $p_\Lambda$ decreasing. For $200 < p_\Lambda < 300$ MeV, the $K^+\Lambda$ correlation function is a little bit larger than one, while for $ p_\Lambda > 300$ MeV, it is slightly smaller than one again. This indicates the repulsive interactions in the $K^+\Lambda$ system at low relative momenta. Indeed, with the fitted parameters, the obtained $K^+ \Lambda$ scattering length is $a_{K^+\Lambda} = 0.33 - i0.14$ and $0.26 - i 0.15$ for Scenario I and II, respectively, which is in agreement with the value obtained in Refs.~\cite{Molina:2023jov,ALICE:2023wjz,ALICE:2020wvi} within errors. But, it is only half the experimental results of Refs.~\cite{ALICE:2023wjz,ALICE:2020wvi}. This discrepancy may be because of the differences in the amplitude used: our multi-coupled-channel framework inherently incorporates intermediate hadronic loops, whereas the experimental analysis adopted a single-channel amplitude parametrization neglecting coupled-channel effects. Furthermore, the theoretical results show a small bump structure near $q \approx 235$ MeV, attributable to coupled-channel effects from $K^+\Sigma^0$ and $K^0\Sigma^+$ channels. It is worthy to mention that the bump structure shown in Scenario II is smoother than the results obtained in Scenario I.

\begin{figure}[htbp]
    \centering
        \includegraphics[scale=0.35]{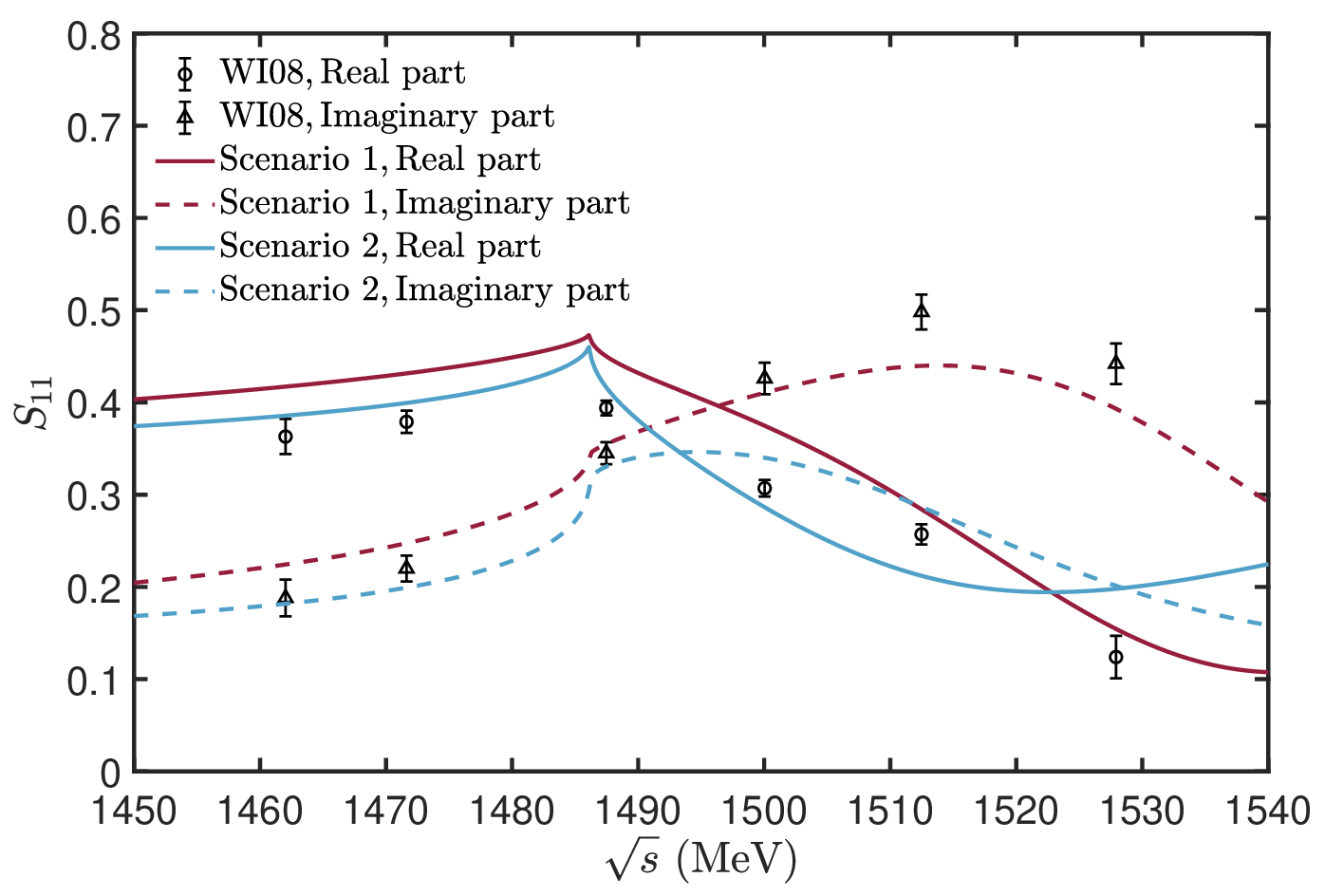}
\caption{Theoretical results for the $\pi N$ elastic scattering amplitude $S_{11}$. The experimental data are taken from Ref.~\cite{Workman:2012hx} of the WI08 solution.}
    \label{fig:pionNscattering}
\end{figure}

Now we can evaluate the two-body $S$-wave $\pi N \to \pi N$ scattering amplitude in the chiral unitary approach with the fitted model parameters. In Fig.~\ref{fig:pionNscattering}, the numerical results of the $\pi N$ scattering amplitudes $S_{11}$ with the fitted parameters of Scenario I and II are shown, compared with the experimental data taken from Ref.~\cite{Workman:2012hx}. One can see that, in the energies around $1450 - 1540$ MeV, the real part of the $\pi N$ scattering amplitude $S_{11}$ can be well reproduced for both Scenario I and II. However, we cannot get a good fit to the imaginary part of the $\pi N$ scattering amplitudes above 1510 MeV for Scenario II.

\section{Summary and Conclusions} \label{sec:Summary and Conclusions}

In this study, we employ the Koonin-Pratt formalism to analyze coupled-channel correlation functions. Focusing on the $K \Lambda$ system, we study the $K^+ \Lambda$ correlation function with coupled-channel $\pi N$, $K\Lambda$, $K \Sigma$ and $\eta N$ $S$-wave interactions, where the $N^*(1535)$ resonance can be dynamically generated in the chiral unitary approach. With the two-body $S$-wave scattering amplitudes obtained within the chiral unitary approach, we then easily get the $K^+ \Lambda$ correlation functions. From fitting our theoretical calculations to the experimental measurements, we can get a good reproduction of the experimental data on the $K^+\Lambda$ correlation function. It is also shown that the $\pi^+ n$ and $\pi^0 p$ channels are crucial to the $K^+ \Lambda$ correlation function, even if their mass thresholds are far from the considered energy region. This work shows the importance of the coupled channel effects in the studying of the correlation functions.

Furthermore, based on the fitted model parameters, we have successfully determined the properties of the $N^*(1535)$ resonance which are also consistent with both the current experimental data and previous theoretical studies within the molecular nature of the state. It is expected that more precise experimental measurements about the correlation functions, including those for other coupled channels, will provide valuable insights into the investigation of exotic states whose fundamental nature remains unknown.

\section*{Acknowledgments}

We would like to thank Profs. Eulogio Oset, Raquel Molina and En Wang for useful discussions and careful reading the manuscript. This work is partly supported by the National Key R\&D Program of China under Grant No. 2023YFA1606703, and by the National Natural Science Foundation of China under Grant Nos. 12435007 and 12361141819.

\bibliography{reference}

\end{document}